\documentclass[prl,reprint,superscriptaddress,footinbib]{revtex4-1}
\usepackage{lipsum}
\usepackage{graphicx}
\usepackage{epsfig}
\pdfoutput=1
\usepackage{subfigure}
\usepackage{xspace}
\usepackage{dcolumn}
\usepackage{multirow}
\usepackage{longtable}
\usepackage{amsmath}
\usepackage{mathtools}
\usepackage{xcolor}
\usepackage[capitalize]{cleveref}
\usepackage{adjustbox}

\begin{document}

\title{Search for Boosted Dark Matter Interacting With Electrons in Super-Kamiokande}

\newcommand{\AFFicrr}{\affiliation{Kamioka Observatory, Institute for Cosmic Ray Research, University of Tokyo, Kamioka, Gifu 506-1205, Japan}}
\newcommand{\AFFkashiwa}{\affiliation{Research Center for Cosmic Neutrinos, Institute for Cosmic Ray Research, University of Tokyo, Kashiwa, Chiba 277-8582, Japan}}
\newcommand{\AFFipmu}{\affiliation{Kavli Institute for the Physics and
Mathematics of the Universe (WPI), The University of Tokyo Institutes for Advanced Study,
University of Tokyo, Kashiwa, Chiba 277-8583, Japan }}
\newcommand{\AFFmad}{\affiliation{Department of Theoretical Physics, University Autonoma Madrid, 28049 Madrid, Spain}}
\newcommand{\AFFubc}{\affiliation{Department of Physics and Astronomy, University of British Columbia, Vancouver, BC, V6T1Z4, Canada}}
\newcommand{\AFFbu}{\affiliation{Department of Physics, Boston University, Boston, MA 02215, USA}}
\newcommand{\AFFbnl}{\affiliation{Physics Department, Brookhaven National Laboratory, Upton, NY 11973, USA}}
\newcommand{\AFFuci}{\affiliation{Department of Physics and Astronomy, University of California, Irvine, Irvine, CA 92697-4575, USA }}
\newcommand{\AFFcsu}{\affiliation{Department of Physics, California State University, Dominguez Hills, Carson, CA 90747, USA}}
\newcommand{\AFFcnm}{\affiliation{Department of Physics, Chonnam National University, Kwangju 500-757, Korea}}
\newcommand{\AFFduke}{\affiliation{Department of Physics, Duke University, Durham NC 27708, USA}}
\newcommand{\AFFfukuoka}{\affiliation{Junior College, Fukuoka Institute of Technology, Fukuoka, Fukuoka 811-0295, Japan}}
\newcommand{\AFFgifu}{\affiliation{Department of Physics, Gifu University, Gifu, Gifu 501-1193, Japan}}
\newcommand{\AFFgist}{\affiliation{GIST College, Gwangju Institute of Science and Technology, Gwangju 500-712, Korea}}
\newcommand{\AFFuh}{\affiliation{Department of Physics and Astronomy, University of Hawaii, Honolulu, HI 96822, USA}}
\newcommand{\AFFicl}{\affiliation{Department of Physics, Imperial College London , London, SW7 2AZ, United Kingdom }}
\newcommand{\AFFkek}{\affiliation{High Energy Accelerator Research Organization (KEK), Tsukuba, Ibaraki 305-0801, Japan }}
\newcommand{\AFFkobe}{\affiliation{Department of Physics, Kobe University, Kobe, Hyogo 657-8501, Japan}}
\newcommand{\AFFkyoto}{\affiliation{Department of Physics, Kyoto University, Kyoto, Kyoto 606-8502, Japan}}
\newcommand{\AFFliv}{\affiliation{Department of Physics, University of Liverpool, Liverpool, L69 7ZE, United Kingdom}}
\newcommand{\AFFmiyagi}{\affiliation{Department of Physics, Miyagi University of Education, Sendai, Miyagi 980-0845, Japan}}
\newcommand{\AFFnagoya}{\affiliation{Institute for Space-Earth Enviromental Research, Nagoya University, Nagoya, Aichi 464-8602, Japan}}
\newcommand{\AFFkmi}{\affiliation{Kobayashi-Maskawa Institute for the Origin of Particles and the Universe, Nagoya University, Nagoya, Aichi 464-8602, Japan}}
\newcommand{\AFFpol}{\affiliation{National Centre For Nuclear Research, 00-681 Warsaw, Poland}}
\newcommand{\AFFsuny}{\affiliation{Department of Physics and Astronomy, State University of New York at Stony Brook, NY 11794-3800, USA}}
\newcommand{\AFFokayama}{\affiliation{Department of Physics, Okayama University, Okayama, Okayama 700-8530, Japan }}
\newcommand{\AFFosaka}{\affiliation{Department of Physics, Osaka University, Toyonaka, Osaka 560-0043, Japan}}
\newcommand{\AFFox}{\affiliation{Department of Physics, Oxford University, Oxford, OX1 3PU, United Kingdom}}
\newcommand{\AFFqmul}{\affiliation{School of Physics and Astronomy, Queen Mary University of London, London, E1 4NS, United Kingdom}}
\newcommand{\AFFregina}{\affiliation{Department of Physics, University of Regina, 3737 Wascana Parkway, Regina, SK, S4SOA2, Canada}}
\newcommand{\AFFseoul}{\affiliation{Department of Physics, Seoul National University, Seoul 151-742, Korea}}
\newcommand{\AFFsheff}{\affiliation{Department of Physics and Astronomy, University of Sheffield, S10 2TN, Sheffield, United Kingdom}}
\newcommand{\AFFshizuokasc}{\affiliation{Department of Informatics in
Social Welfare, Shizuoka University of Welfare, Yaizu, Shizuoka, 425-8611, Japan}}
\newcommand{\AFFstfc}{\affiliation{STFC, Rutherford Appleton Laboratory, Harwell Oxford, and Daresbury Laboratory, Warrington, OX11 0QX, United Kingdom}}
\newcommand{\AFFskk}{\affiliation{Department of Physics, Sungkyunkwan University, Suwon 440-746, Korea}}
\newcommand{\AFFtokyo}{\affiliation{The University of Tokyo, Bunkyo, Tokyo 113-0033, Japan }}
\newcommand{\AFFtodai}{\affiliation{Department of Physics, University of Tokyo, Bunkyo, Tokyo 113-0033, Japan }}
\newcommand{\AFFtit}{\affiliation{Department of Physics,Tokyo Institute of Technology, Meguro, Tokyo 152-8551, Japan }}
\newcommand{\AFFtus}{\affiliation{Department of Physics, Faculty of Science and Technology, Tokyo University of Science, Noda, Chiba 278-8510, Japan }}
\newcommand{\AFFtoronto}{\affiliation{Department of Physics, University of Toronto, ON, M5S 1A7, Canada }}
\newcommand{\AFFtriumf}{\affiliation{TRIUMF, 4004 Wesbrook Mall, Vancouver, BC, V6T2A3, Canada }}
\newcommand{\AFFtokai}{\affiliation{Department of Physics, Tokai University, Hiratsuka, Kanagawa 259-1292, Japan}}
\newcommand{\AFFtsinghua}{\affiliation{Department of Engineering Physics, Tsinghua University, Beijing, 100084, China}}
\newcommand{\AFFuw}{\affiliation{Department of Physics, University of Washington, Seattle, WA 98195-1560, USA}}
\newcommand{\AFFynu}{\affiliation{Faculty of Engineering, Yokohama National University, Yokohama, 240-8501, Japan}}
\AFFicrr
\AFFkashiwa
\AFFmad
\AFFbu
\AFFubc
\AFFbnl
\AFFuci
\AFFcsu
\AFFcnm
\AFFduke
\AFFfukuoka
\AFFgifu
\AFFgist
\AFFuh
\AFFicl
\AFFkek
\AFFkobe
\AFFkyoto
\AFFliv
\AFFmiyagi
\AFFnagoya
\AFFkmi
\AFFpol
\AFFsuny
\AFFokayama
\AFFosaka
\AFFox
\AFFqmul
\AFFregina
\AFFseoul
\AFFsheff
\AFFshizuokasc
\AFFstfc
\AFFskk
\AFFtokai
\AFFtokyo
\AFFtodai
\AFFipmu
\AFFtit
\AFFtus
\AFFtoronto
\AFFtriumf
\AFFtsinghua
\AFFuw
\AFFynu

\author{C.~Kachulis}
\AFFbu

\author{K.~Abe}
\AFFicrr
\AFFipmu
\author{C.~Bronner}
\AFFicrr
\author{Y.~Hayato}
\AFFicrr
\AFFipmu
\author{M.~Ikeda}
\AFFicrr
\author{K.~Iyogi}
\AFFicrr 
\author{J.~Kameda}
\AFFicrr
\AFFipmu 
\author{Y.~Kato}
\AFFicrr
\author{Y.~Kishimoto}
\AFFicrr
\AFFipmu 
\author{Ll.~Marti}
\AFFicrr
% ...


\collaboration{The Super-Kamiokande Collaboration}
\noaffiliation

\date{\today}

\begin{abstract}
A search  for boosted dark matter using 161.9 kiloton-years of Super-Kamiokande IV data is presented.  We search for an excess of elastically scattered electrons above the atmospheric neutrino background, with a visible energy between 100 MeV and 1 TeV, pointing back to the Galactic Center or the Sun.  No such excess is observed.  Limits on boosted dark matter event rates in multiple angular cones around the Galactic Center and Sun are calculated.  Limits are also calculated for a baseline model of boosted dark matter produced from cold dark matter annihilation or decay.  This is the first experimental search for boosted dark matter from the Galactic Center or the Sun interacting in a terrestrial detector.  
\end{abstract}
\maketitle

While there has long been ample evidence for the existence of dark matter \cite{Zwicky:1933gu,Rubin:1980dm,Blumenthal:1984eu,Begeman:1991iy,Bertone:2004pz}, the specific properties and identity of dark matter remain elusive.  The $\Lambda$CDM cosmology, which consists of long lived dark matter that was non-relativistic (``cold") at freeze-out and a cosmological constant $\Lambda$, which corresponds to dark energy, has been well supported by cosmological observations \cite{Bamba:2012cp}.  Under this cosmology, the dark matter abundance has been measured by observation of the Cosmic Microwave Background (CMB) to account for about 25\% of the energy density of the Universe \cite{Ade:2015xua,Hinshaw:2012aka}.  However, despite numerous direct and indirect detection searches, as well as searches for dark matter produced at particle accelerators, there has thus far been no definitive observation of particle dark matter \cite{Akerib:2013tjd,Ahmed:2009zw,Aprile:2012nq,Desai:2004pq,Aartsen:2017ulx,Adrian-Martinez:2013ayv,Khachatryan:2014rra,Aad:2015zva,Goodman:2010ku}.\par
With the properties of dark matter so uncertain, various possibilities must be considered.  One possibility is that some dark matter is in fact not cold, but is highly relativistic and has been produced at late times, thus denoted ``boosted" dark matter \cite{Agashe:2014yua,Huang:2013xfa,Cherry:2015gw,Bhattacharya:2014yha,Kopp:2015gp,Kong:2015jb,Berger:2014hq,Kim:2016zjx}. Boosted dark matter could exist as a subdominant dark matter component, with a dominant cold dark matter component accounting for most of the dark matter energy density of the Universe.  In this way, boosted dark matter can remain consistent with $\Lambda$CDM.  The subdominant boosted dark matter can be the same particle as the dominant cold dark matter, or it can be a different, lighter particle.  Boosted dark matter can be produced from the dominant cold dark matter through a variety of processes, including annihilation \cite{Belanger:2011ww,SungCheon:2008ts}, semi-annihilation \cite{DEramo:2010keq,Hambye:2008bq,Hambye:2009fg,Arina:2009uq,Belanger:2012vp}, number-changing 3$\rightarrow$2 self-annihilation \cite{Carlson:1992fn,deLaix:1995vi,Hochberg:2014dra}, and decay \cite{Kopp:2015gp,Bhattacharya:2016tma}.  Boosted dark matter can then be observed through its scattering off electrons or nuclei in large volume terrestrial detectors \cite{Necib:2016aez,Alhazmi:2016qcs}.  Current direct detection limits can be evaded in multi-component models by having only the boosted dark matter species couple directly to Standard Model particles \cite{Agashe:2014yua,Kopp:2015gp,Kong:2015jb,Berger:2014hq} or in boosted dark matter single-component models by invoking a spin dependent dark matter-nucleon cross section \cite{Berger:2014hq}. \par
This letter reports the results of a search for boosted dark matter coupling to electrons in Super-Kamiokande (SK), with the boosted dark matter originating in the Galactic Center or the Sun, and with scattered electron energies ranging from 100 MeV to 1 TeV.  This is the first time that this class of high energy ``electron elastic scatter-like" events has been studied at SK.  The search is performed on 2,628.1 days of SK-IV data, which corresponds to 161.9 kiloton-years (kT-y) exposure.  The analysis is designed to be independent of the particular model of the coupling between boosted dark matter and electrons.  This way, the results can be applied to any model that predicts a source of particles from the Galactic Center or Sun which would scatter electrons to energies greater than 100 MeV. \par
The SK detector \cite{Fukuda:2002uc} is a water Cherenkov detector located 1,000 meters below Mt. Ikenoyama in Gifu, Japan.  It consists of a 50 kT cylindrical tank of water, which is divided into a 32 kT (22.5 kT fiducial) inner detector (ID) surrounded by an outer detector (OD).  The ID and the OD are optically separated by black Tyvek sheeting, and the ID is observed by 11,129 inward-facing 20-inch photo multiplier tubes (PMTs), while the OD is observed by 1,885 outward-facing 8-inch PMTs.  The ID provides most of the information used in event reconstruction, while the OD is used as an active veto region.  Events are classified as Fully-Contained (FC) if there is activity only in the ID.  Relativistic charged particles in SK produce Cherenkov rings, which are categorized as $e$-like for electrons and $\gamma$'s, or $\mu$-like for muons. \par
The search begins with the Fully-Contained Fiducial-Volume (FCFV) dataset.  This dataset  consists of all  events with no OD activity, reconstructed vertex inside the fiducial volume, and greater than 100 MeV of $e$-like momentum.  These are events that originate and deposit all of their energy in the ID.  This is a standard SK dataset used to study atmospheric neutrinos \cite{Wendell:2010md}.  From this dataset, we search for elastically scattered electrons by applying the following analysis cuts: \\
\\
\indent 
1. 1-ring (if E$_{vis}<$100 GeV)\\
\indent
2. $e$-like\\
\indent
3. 0 decay electrons\\
\indent
4. 0 tagged neutrons\\
\\
The first two cuts search for a single relativistic electron, while the final two cuts remove events with a signature of a nuclear interaction.  Decay electrons in $e$-like events are the result of the $\pi^\pm \rightarrow \mu^\pm \rightarrow e^\pm$ decay chain with the $\pi^\pm$ coming from a neutrino-nucleus interaction.  Tagged neutrons originate from neutrons being knocked out of the nucleus following a neutrino-nucleus interaction, thermalizing, and capturing on Hydrogen.  Neutron captures are particularly numerous following neutrino deep inelastic scattering.  Neither decay electrons nor neutron captures should occur following the elastic scatter of an electron by a boosted dark matter particle.  The 1-ring cut is not applied for events with visible energy above 100 GeV, as the ring counting algorithm, which is tuned for lower energy events, becomes unreliable at such high energies.  We choose to restrict this analysis to SK-IV data only in order to take advantage of neutron tagging to remove atmospheric neutrino background.  While neutron tagging has been used similarly in SK proton decay searches \cite{Miura:2016krn}, this is the first time that it has been applied toward sample purification at energies greater than about 1 GeV.  Event displays of example events passing the analysis cuts are provided in the Supplemental Material \footnote{See Supplemental Material at [URL] for event displays of example events passing the analysis cuts}. \par
Due to intricacies of the SK-IV trigger logic, events $\gtrsim 50$ GeV often cannot have neutron tagging applied.  Events for which neutron tagging cannot be applied are considered signal if they pass the first three analysis cuts.  Fourteen of such events were found by FCFV selection, none of which passed the first three analysis cuts.  This effect is accounted for in the signal efficiency, background rate, and limit calculations described below.  The effect is minimal, since the efficiency of the neutron tagging cut is nearly 100\%, and the background rate is driven down at such high energies by the sharp drop-off of the atmospheric neutrino flux. \par
\begin{figure}
	\includegraphics[width=0.45\textwidth,keepaspectratio]{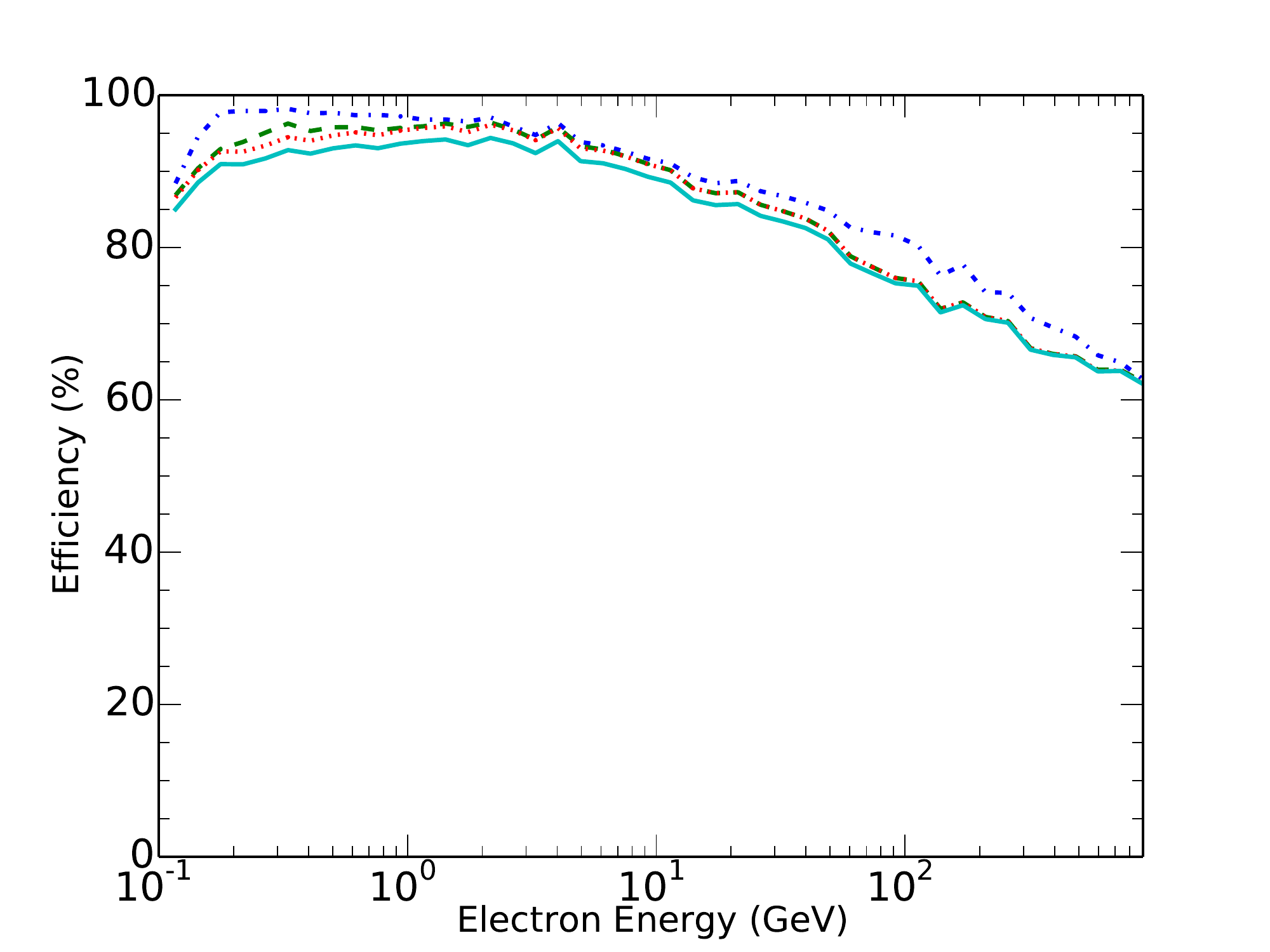}
	\caption{Signal efficiency of the FCFV selection and analysis cuts as a function of energy.  Beginning with the FCFV selection (dashed-dotted blue), the addition of the 1-ring (for E$_{vis}<$100GeV, dashed green), $e$-like (dotted red) and finally 0 decay electrons and 0 tagged neutrons cuts to arrive at the final efficiency (solid cyan) are shown.  The efficiency of the 0 decay electrons cut is $>99.99\%$, so that the drop from the dotted red line to solid cyan line is due solely to the neutron tagging cut.}
	\label{fig:eff}
\end{figure}

The efficiency of the analysis cuts was found using a signal MC of 200,000 electrons with energies ranging from 30 MeV to 1 TeV.  Events were simulated up to 1 meter outside the fiducial volume, and efficiency was defined as the number of events passing each cut divided by the number of events simulated in the fiducial volume (events can migrate into or out of the fiducial volume during reconstruction).   The efficiency of the FCFV selection and analysis cuts is shown in \cref{fig:eff}.  The main cause of the reduction of efficiency with energy is the loss of containment at high energies; some higher energy electromagnetic showers are able to penetrate from the FV into the OD, and so do not pass the FC selection.  The systematic uncertainty on the efficiency is estimated to be about 2\%, most of which is due to uncertainty in the efficiency of the 1-ring cut.      The small size of the uncertainty of the FCFV selection and first three cuts is due to significant separation between passing and failing events, so that very few signal events occur near the boundaries of the cuts.  The efficiency of the neutron tagging cut is found by measuring the rate of false neutron tags in real data recorded by a periodic trigger, and thus has minimal systematic uncertainty. \par
 Because the atmospheric neutrino background to this search is strongly energy dependent, events are separated into three samples based on visible energy with ranges 100 MeV$<$E$_{\textrm{vis}}<$1.33 GeV, 1.33 GeV$<$E$_{\textrm{vis}}<$20 GeV , and E$_{\textrm{vis}}>$20 GeV. The number of data events is shown for each sample in \cref{tab:cut_counts}, along with the simulated atmospheric neutrino Monte-Carlo (MC) expectation. The signal efficiency at representative energies based on signal electron MC is also shown.  The importance of the decay electron and neutron tagging cuts is particularly evident in the highest energy sample ($E_{vis}>20$ GeV), where they together reduce the background by about a factor of 10 while having a minimal effect on signal efficiency.\par
Since boosted dark matter is expected to originate in regions of high dark matter density, this search looks for a signal coming from the Galactic Center or the Sun (some boosted dark matter models predict a significant capture rate of cold dark matter in the Sun, either through a spin dependent dark matter-nucleus cross section \cite{Cherry:2015gw} or through the combination of a relatively strong dark matter self interaction and coupling between cold dark matter and Standard Model particles through boosted dark matter loops \cite{Kong:2015jb}).  Cones are drawn around the signal source, and the number of events passing the analysis cuts in each cone is counted.  While the exact relationship between the boosted dark matter and the scattered electron directions is model dependent, the scattering is in general expected to be strongly forward since the energy of the boosted dark matter and the electron recoil energy are both assumed to be much greater than the mass of the electron.  Under these assumptions, the scattering angle of the recoil electron is kinematically constrained to be less than $5.8^\circ$ at a recoil energy of 100 MeV, with the maximum allowed scattering angle decreasing as $1/\sqrt{E_e}$. Thus, since the reconstructed electron direction was found to have an angular resolution better than $3^\circ$ over the entire energy range, it is a good proxy for the direction of the boosted dark matter.  \par
When the source of the signal is the Galactic Center, the optimal size of the search cone is dependent on both the distribution of the dominant dark matter species in the Galaxy, and the production method of the boosted dark matter.  Production of boosted dark matter through both dark matter annihilation and decay were considered for three dark matter halo models: Moore \cite{Moore:1999kx}, NFW \cite{Navarro:1996ce}, and Kravtsov \cite{Kravtsov:1998it}.  For each combination of halo model and production method, the signal MC was reweighted assuming the direction of the scattered electron was the same as the direction of the boosted dark matter.  Optimal cone angles were found by maximizing Efficiency/$\sqrt{\textrm{Background}}$.  The optimal half-opening angle of the search cone was found to range from less than $5^\circ$ to around $40^\circ$, depending on halo model and boosted dark matter production method. We therefore used eight search cones around the Galactic Center, ranging from $5^\circ$ to $40^\circ$ in steps of $5^\circ$.  When the Sun is the signal source the situation is much simpler, since it is effectively a point source.  Therefore, a single search cone of 5$^\circ$ around the Sun was used for the solar search.\par
 
\begin{table*}
\begin{ruledtabular}
\begin{tabular}{lccc@{\hspace{.3in}}ccc@{\hspace{.3in}}ccc}
& \multicolumn{3}{c}{100 MeV$<E_{vis}<1.33$ GeV} & \multicolumn{3}{c}{1.33 GeV $<E_{vis}<20$ GeV} & \multicolumn{3}{c}{$E_{vis}>$ 20 GeV}\\
&Data&$\nu$-MC&$\epsilon_{sig}$(0.5 GeV)&Data&$\nu$-MC&$\epsilon_{sig}$(5 GeV)&Data&$\nu$-MC&$\epsilon_{sig}$(50 GeV)\\
\hline
FCFV & 15206 & 14858.1&97.7\%& 4908 & 5109.7&93.8\% & 118 & 107.5&84.9\% \\
\& single ring & 11367& 10997.4 &95.8\%& 2868& 3161.8&93.3\% & 71& 68.2&82.2\% \\
\& $e$-like & 5655& 5571.5 &94.7\%& 1514& 1644.2&93.0\% & 71 & 68.1&82.2\%\\
\& 0 decay-e & 5049& 5013.8 &94.7\%& 1065& 1207.2&93.0\% & 13 & 15.7&82.2\%\\
\& 0 neutrons & 4042& 3992.9 &93.0\%& 658& 772.6&91.3\% & 3& 7.4&81.1\% \\
\end{tabular}
\end{ruledtabular}
\caption{Number of events over the entire sky passing each cut in 2628.1 days of SK4 data, simulated $\nu$-MC background expectation, and signal efficiency at representative energy after each cut.}
\label{tab:cut_counts}
\end{table*}
A data-driven off-source method was used to estimate background due to atmospheric neutrinos for the two lower energy samples \cite{Li:1983fv}.  In order to avoid contamination from a potential signal, the off-source region is defined as everything outside an $80^\circ$ cone for the Galactic Center search, and everything outside of the $5^\circ$ search cone for the solar search.  The off-source regions are defined, like the search cones, in celestial coordinates for the Galactic Center search, and solar coordinates for the Sun search.  For a particular search cone, each data event in the off-source region can be assigned two values based on its direction in horizontal coordinates $\hat{d}$: $T_{off}$, which is the fraction of time $\hat{d}$ spends within the off-source region, and $T_{on}$, which is the fraction of time $\hat{d}$ spends within the search cone.  The event is then weighted by the ratio $T_{on}/T_{off}$.  The sum of these weights gives an estimate of the background in the search cone, while the square-root of the sum of the squares of the weights gives the uncertainty on this estimate.  The resulting estimates are independent of MC, and their uncertainties range from 3\% to 19\% , which are smaller than the uncertainties on the corresponding estimates based on MC.  \par
\begin{figure}
\includegraphics[width=0.45\textwidth,keepaspectratio]{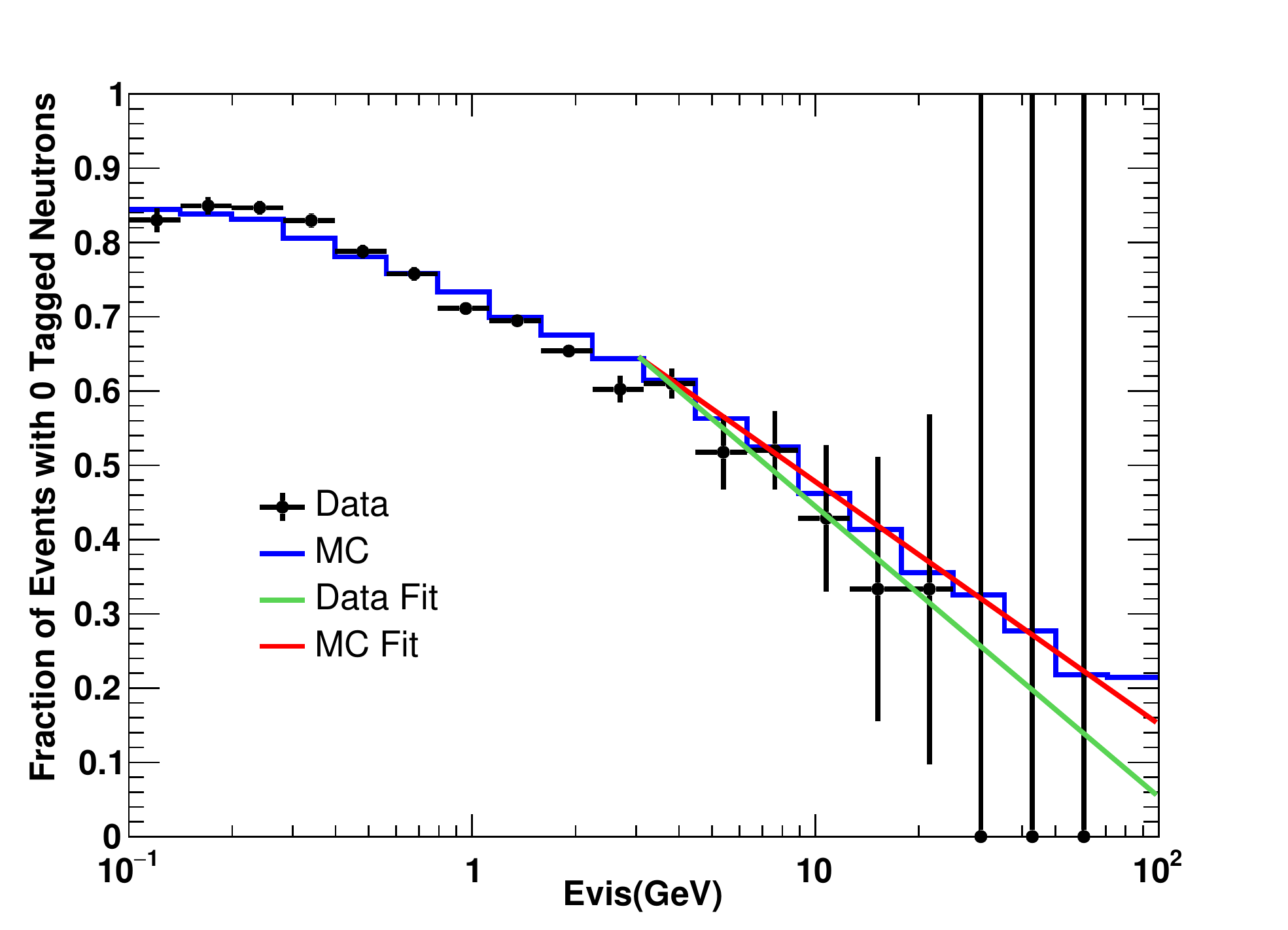}
\caption{Neutron tagging cut systematic error estimation.  Data and MC fits are to logarithmic function.}
\label{fig:ntag_syst}
\end{figure}

While the off-source method works well for the two lower energy samples, there are too few events in the highest energy sample ($E_{vis}>$20 GeV) for it to be successfully applied.  Therefore, the background in the highest energy sample was estimated using the 500-year SK atmospheric neutrino MC.  The MC was livetime normalized and oscillated according to 3-flavor oscillations with nominal oscillation parameters.  The systematic uncertainty on the estimated background was found by summing in quadrature the effects of 1$\sigma$ shifts of 75 systematics \cite{Wendell:2010md}.  The uncertainties in the values of oscillation parameters were included as systematics.   The total uncertainty in the background was estimated to be 30\%. The dominant systematic is the uncertainty related to the neutron tagging cut, which is estimated to be 23\%.  This systematic accounts for the uncertainty in the efficiency of the neutron tagging algorithm, as well as in the modeling of production and transport of neutrons in the detector.  It was estimated using a Data-MC comparison of the fraction of events (over the entire sky) passing the first three analysis cuts that also had zero tagged neutrons.  This comparison is shown in \cref{fig:ntag_syst}.  Above about 10 GeV of visible energy, there are very few events in the data, making a Data-MC comparison difficult.  To compensate, both the data and MC were fit to logarithmic functions $A+B\log \frac{\textrm{E}_\textrm{vis}}{\textrm{GeV}}$, in the region above 3 GeV.  The systematic uncertainty as a function of energy was then taken as the difference between the Data fit and the MC fit.  The value of the neutron tagging systematic for the background estimate of the highest energy sample was found by applying this shift on an event by event basis as a function of the visible energy of the particular event.  While this estimate is rather imprecise, the expected background rates in the signal cones above 20 GeV are all below one, meaning that the systematic uncertainty on the background estimate is minimal compared to fluctuations associated with Poisson counting statistics.  It is thus only the magnitude of this systematic that is important; its exact value has minimal influence on the calculated limits.   \par
The results of the search are shown in \cref{tab:results}. The observed data are consistent with expected background for both the Galactic Center and Sun searches.  In the highest energy sample, the search is essentially background free, and no candidates were found in any of the search cones.  Skymaps of the locations of every event passing the analysis cuts are provided for each energy sample in the Supplemental Material \footnote{See Supplemental Material at [URL] for skymaps of events passing analysis cuts}.
\begin{table*}
\centering
	\begin{ruledtabular}
	\begin{tabular}{lccc@{\hspace{.2in}}ccc@{\hspace{.2in}}ccc}
         & \multicolumn{3}{c}{100 MeV$<E_{vis}<1.33$ GeV} & \multicolumn{3}{c}{1.33 GeV $<E_{vis}<20$ GeV} & \multicolumn{3}{c}{$E_{vis}>$ 20 GeV}\\
	\parbox{2cm}{\raggedright Search \\ Cone}&\parbox[t]{2cm}{Expected \\ Bckg} & Data & \parbox[t]{2cm}{Sig Rate Limit \\ (kT-y)$^{-1}$} &\parbox[t]{2cm}{Expected \\ Bckg} & Data &\parbox[t]{2cm}{Sig Rate Limit \\ (kT-y)$^{-1}$}&\parbox[t]{2cm}{Expected \\ Bckg} & Data&\parbox[t]{2cm}{Sig Rate Limit \\ (kT-y)$^{-1}$}\\
	\hline 
	GC $5^\circ$  &$8.4\pm0.7$ & 5&0.017&$1.6\pm0.3$&1&0.018&$0.016\pm0.005$&0&0.015 \\
	GC $10^\circ$  &$32.0\pm1.9$&24&0.023&$6.3\pm0.84$&5&0.026&$0.060\pm0.018$&0&0.015 \\
	GC $15^\circ$  & $72.5\pm3.5$&69&0.078&$13.6\pm1.6$&11&0.032&$0.14\pm0.04$&0&0.014\\
	GC $20^\circ$ &$126.5\pm5.4$&125&0.123&$23.3\pm2.3$&18&0.028&$0.25\pm0.07$&0&0.014\\
	GC $25^\circ$  &$196.8\pm7.6$&202&0.201& $35.4\pm3.3$&31&0.049&$0.37\pm0.11$&0&0.013\\
	GC $30^\circ$  &$283.7\pm10.1$&285&0.214&$49.3\pm4.3$&48&0.081&$0.53\pm0.16$&0&0.012\\
	GC $35^\circ$  &$384.8\pm12.8$&375&0.187&$68.1\pm5.4$&67&0.101&$0.70\pm0.21$&0&0.011\\
	GC $40^\circ$ &$499.6\pm15.9$&494&0.249&$90.2\pm6.9$&90&0.124&$0.90\pm0.27$&0&0.011\\	
	\hline
	Sun $5^\circ$ &$7.59\pm0.18$&5&0.017&$1.25\pm0.07$&1&0.020&$0.015\pm0.004$&0&0.015\\
	\end{tabular}
         \end{ruledtabular}	
	\caption{Estimated backgrounds,numbers of events in data, and signal event rate limits for each cone and each energy sample.  The event rate limits are at the 90\% confidence level.}
	\label{tab:results}
\end{table*}
For each cone and energy sample, confidence intervals for the observed boosted dark matter event rate were computed using a Poisson $\chi^2$ statistic that incorporates the systematic uncertainty on the background estimate through the pull method \cite{Fogli:2002do,Olive:2016xmw}:
\begin{equation}
\chi^2(s)=\min \limits_\delta\left[ 2  \left(  E -\mathcal{O}  + \mathcal{O} \ln \frac{ \mathcal{O} }{ E } \right)
             + \delta^2\right],
\label{eq:chi2}
\end{equation}
where $E=b(1+\delta\sigma)+s$, $b$ is the estimated number of background events with systematic uncertainty $\sigma$, $s$ is the number of signal events being tested, $\delta$ is the systematic pull that is minimized over, and $\mathcal{O}$ is the observed number of events.  Note that for the two lower energy samples, the background systematic uncertainty $\sigma$ is due to statistical fluctuations associated with the off-source method, while for the highest energy sample it is due to uncertainties in the MC production.  The test statistic $\Delta \chi^2$ was calculated by subtracting the global minimum $\chi^2$.  To find the confidence level at which a particular value of $s$ is allowed, the measured value of $\Delta \chi^2$ at that value of $s$ was compared to the $\Delta \chi^2$ distribution of a large number of toy MC produced assuming that level of signal.  Since the $s=0$ hypothesis is allowed at 90\% confidence for all search cones and energy samples, the upper ends of the 90\% confidence intervals are interpreted as 90\% upper limits, and presented in \cref{tab:results}.  \par
To demonstrate the application of this result to a specific model, limits were calculated on a baseline boosted dark matter model with the Galactic Center as the signal source \cite{Agashe:2014yua}.  This model introduces two dark fermions $\psi_A$ and $\psi_B$ and a massive dark photon $\gamma'$, with an assumed mass ordering $m_A>m_B>m_{\gamma'}$.   The particle $\psi_A$ is proposed to be the dominant cold dark matter in the universe, and does not couple directly to Standard Model particles.  The particle $\psi_B$ is the boosted dark matter, and couples to Standard Model particles through the exchange of the dark photon $\gamma '$.  The coupling between $\gamma '$ and $\psi_B$ is set by a coupling constant $g'$ which is proposed to be large but perturbative, while the coupling between $\gamma '$ and $e^-$ is scaled from $\gamma-e^-$ coupling by the constant $\varepsilon$.  Limits were calculated for two scenarios of $\psi_B$ production, one where $\psi_B$ is produced through annihilation of $\psi_A$ with $\bar{\psi}_A$, and another where $\psi_B$ is produced through the decay of $\psi_A$.  In the annihilation scenario, the thermal annihilation cross section is set to $\langle \sigma_{A\bar{A} \rightarrow B\bar{B}} v\rangle=5 \times 10^{-26}$ cm$^3$/s in order to achieve the observed relic density $\Omega_A \approx 0.2$ through an assisted freeze out scenario \cite{Belanger:2011ww,Agashe:2014yua}.  The energy of $\psi_B$ is equal to $m_A$ in this scenario.  In the decay scenario, the decay lifetime of $\psi_A$, $\tau_{decay}$, is taken to be a free parameter, and the energy of $\psi_B$ is assumed to be $m_A/2$.\par
\begin{figure}
\includegraphics[width=0.45\textwidth, keepaspectratio]{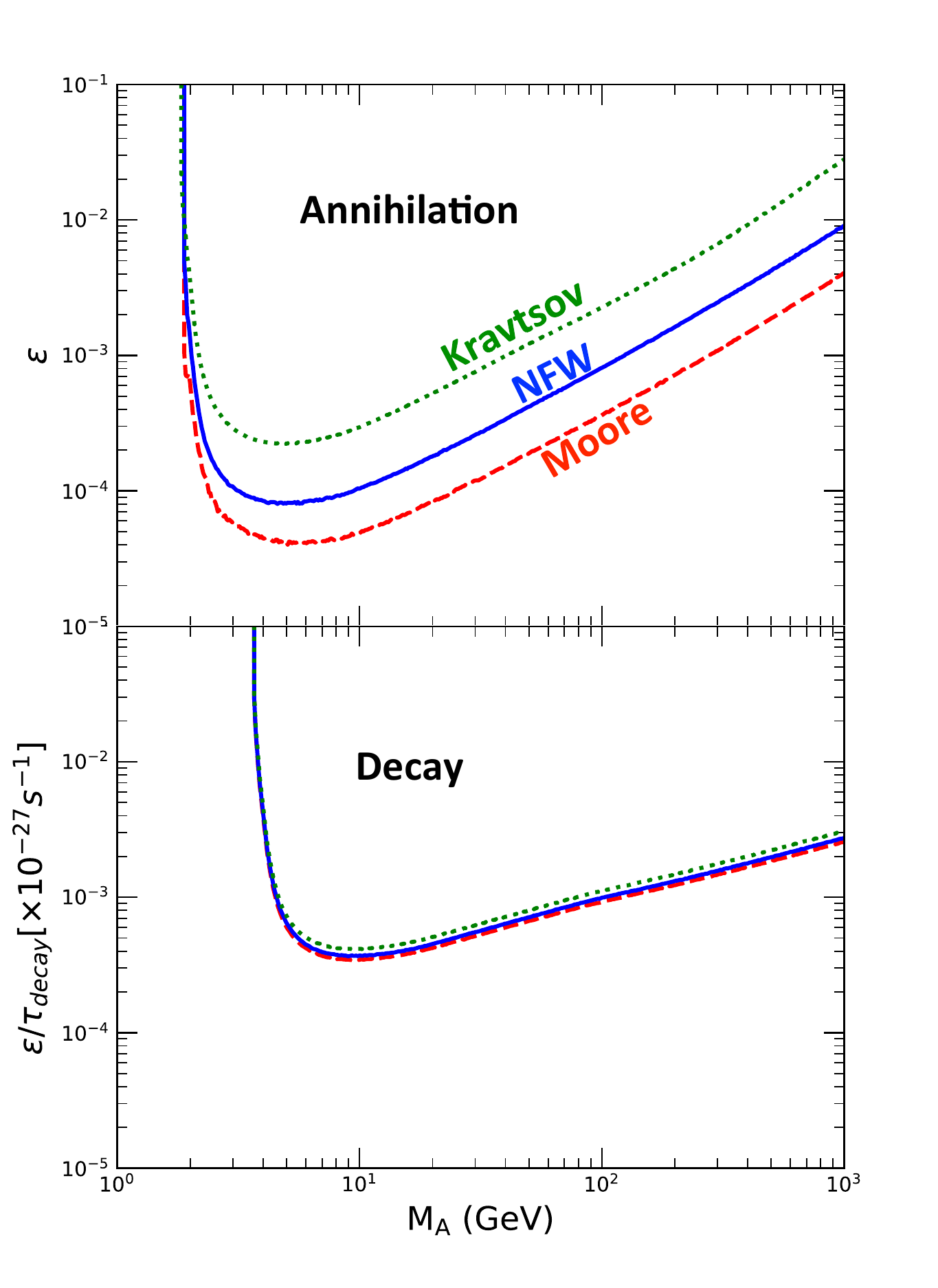}
\caption{90\% Confidence Interval upper limits for $m_B$=200 MeV, $m_\gamma '$=20 MeV, and $g'$=0.5, for boosted dark matter produced by annihilation (top) and decay (bottom). }
\label{fig:limits_mA_epsilon}
\end{figure}

Limits were calculated separately for the Moore, NFW, and Kravtsov Galactic halo models, using the results from a different cone for each fit.  For the annihilation scenario, the $5^\circ$ cone was used for the Moore model, the $10^\circ$ cone for the NFW model, and the $40^\circ$ cone for the Kravtsov model.  For the decay scenario, the $40^\circ$ cone was used for all three galactic halo models.  These cones were selected using the cone optimization technique described earlier. For each halo model, signal MC events were reweighted based on the values of $m_A, m_B, m_{\gamma '}, \varepsilon$ and $g'$ at the particular point in parameter space being tested.  This reweighting accounts for the model dependent recoil electron energy spectrum, as well as the model dependent smearing between the boosted dark matter direction and the recoil electron direction.  The effect of boosted dark matter scattering off of both electrons and protons in the Earth is also accounted for, though this effect is negligible for the majority of the allowed parameter space.  A binned $\chi^2$ statistic was then computed similar to the one described above:
\begin{equation}
\chi^2=\sum \limits_i^3 \min \limits_{\delta_i}\left[ 2  \left(  E_i-\mathcal{O}_i  + \mathcal{O}_i \ln \frac{ \mathcal{O}_i }{ E_i } \right)
             + \delta_i^2\right],
\end{equation}
with variables defined as before, summed over three bins corresponding to the three energy samples.  The $\Delta \chi^2$ test statistic was then calculated by subtracting the global minimum $\chi^2$.  Confidence intervals were found by comparing the measured $\Delta \chi^2$ values with the distributions of $\Delta \chi^2$ values found by many toy Monte Carlo simulations produced at each point.  Ninety-percent confidence intervals were computed in the $\varepsilon$ vs $m_A$ plane for the annihilation scenario, and the $\varepsilon/\tau_{decay}$ vs $m_A$ plane for the decay scenario, with $m_B$, $m_\gamma '$ and $g'$ set to representative values of $m_B$=200 MeV, $m_\gamma '$=20 MeV, and $g'$=0.5.  Since the $\varepsilon=0$ points, which correspond to no signal, are allowed at 90\% confidence, the resulting confidence intervals are interpreted as upper limits.  These limits are shown for the Moore, NFW, and Kravtsov halo models in \cref{fig:limits_mA_epsilon}. \par
In summary, we have searched for evidence of boosted dark matter by looking for high energy elastically scattered electrons that point back to the Galactic Center or the Sun.  We have found no such evidence.  This is the first study of high energy ``electron elastic scatter-like" events at SK.  The use of decay electron and tagged neutron cuts significantly reduced background in the highest energy sample, allowing for an effectively background free search in that energy range.  Our results are presented in a model independent way, which makes them applicable not only to boosted dark matter, but to any theory that predicts an excess of particles from the Galactic Center or Sun that would elastically scatter electrons to energies above 100 MeV.  
\\
\begin{acknowledgements}
We gratefully acknowledge the cooperation of the Kamioka Mining and Smelting Company. The Super-‐Kamiokande experiment has been built and operated from funding by the Japanese Ministry of Education, Culture, Sports, Science and Technology, the U.S. Department of Energy, and the U.S. National Science Foundation. Some of us have been supported by funds from the National Research Foundation of Korea NRF-‐2009-‐0083526 (KNRC) funded by the Ministry of Science, ICT, and Future Planning, the European Union H2020 RISE-‐GA641540-‐SKPLUS, the Japan Society for the Promotion of Science, the National Natural Science Foundation of China under Grants No. 11235006, the National Science and Engineering Research Council (NSERC) of Canada, the Scinet and Westgrid consortia of Compute Canada, and the National Science Centre, Poland (2015/17/N/ST2/04064, 2015/18/E/ST2/00758).
\end{acknowledgements}
\bibliographystyle{apsrev4-1}
%

\end{document}